\documentclass[12pt]{article}
\begin{document}
\title{Neutron Stars and Fractal Dimensionality}
\author{B.G. Sidharth\\
International Institute for Applicable Mathematics \& Information Sciences\\
Hyderabad (India) \& Udine (Italy)\\
B.M. Birla Science Centre, Adarsh Nagar, Hyderabad - 500 063
(India)}
\date{}
\maketitle
\begin{abstract}
We argue that the material inside Neutron stars behaves anomalously
with fractal statistics and that in principle, we could induce mini
Neutron stars, with the release of energy.
\end{abstract}
\section{Introduction}
It is known that in certain special cases Fermions and Bosons obey
what may be called Fractal Statistics, that is Fermions exhibit
bosonization effects and vice versa while at the same time the space
dimensionality appears to be less than $3$ \cite{bgscsf1,bgsjsp,cu}.
We first recapitulate some of these
results.\\
Let us consider a nearly mono energetic collection of Fermions. We
next use the well known formula of the occupation number of a Fermi
gas \cite{huang}
\begin{equation}
\bar n_p = \frac{1}{z^{-1}e^{bE_p}+1}\label{8je9}
\end{equation}
where, $z' \equiv \frac{\lambda^3}{v} \equiv \mu z \approx z$
because, here, as can be easily shown $\mu \approx 1,$
$$v = \frac{V}{N}, \lambda = \sqrt{\frac{2\pi \hbar^2}{m/b}}$$
\begin{equation}
b \equiv \left(\frac{1}{KT}\right), \quad \mbox{and} \quad \sum \bar
n_p = N\label{8je10}
\end{equation}
Let us consider in particular a collection of Fermions which is
somehow made nearly mono-energetic, that is, given by the
distribution,
\begin{equation}
n'_p = \delta (p - p_0)\bar n_p\label{8je11}
\end{equation}
where $\bar n_p$ is given by (\ref{8je9}).\\
This is not possible in general - here we consider a special
situation of a collection of mono-energetic particles in equilibrium
which is the idealization
of a contrived experimental set up.\\
By the usual formulation we have,
\begin{equation}
N = \frac{V}{\hbar^3} \int d\vec p n'_p = \frac{V}{\hbar^3} \int
\delta (p - p_0) 4\pi p^2\bar n_p dp = \frac{4\pi V}{\hbar^3} p^2_0
\frac{1}{z^{-1}e^{\theta}+1}\label{8je12}
\end{equation}
where $\theta \equiv bE_{p_0}$.\\
It must be noted that in (\ref{8je12}) there is a loss of dimension
in momentum space, due to the $\delta$ function in (\ref{8je11}) -
in fact such a fractal two dimensional situation would in the
relativistic case lead us back to the anomalous behaviour already
alluded to (Cf.ref.\cite{bgscosmic} for details). In the non
relativistic case two dimensions would imply that the coordinate
$\psi$ of the spherical polar coordinates $(r, \psi , \phi)$ would
become constant, $\pi /2$ in fact. In this case the usual Quantum
numbers $l$ and $m$ of the spherical harmonics \cite{powell} no
longer play a role in the usual radial wave equation
\begin{equation}
\frac{d^2u}{dr^2} + \left\{\frac{2m}{\hbar^2}[E - V(r)] -
\frac{l(l+1)}{r^2}\right\} u = 0,\label{8je13}
\end{equation}
The coefficient of the centrifugal term $l(l+1)$ in (\ref{8je13}) is
replaced by $m^2$ as in
Classical Theory\cite{gold}.\\
To proceed, in this case, $KT = <E_p> \approx E_{p}$ so that,
$\theta \approx 1$.
But we can continue without giving $\theta$ any specific value.\\
Using the expressions for $v$ and $z$ given in (\ref{8je10}) in
(\ref{8je11}), we get
$$(z^{-1} e^\theta + 1) = (4\pi )^{5/2} \frac{z^{'-1}}{p_0};\mbox{whence}$$
\begin{equation}
z^{'-1}A\equiv z^{'-1}\left(\frac{(4\pi )^{5/2}}{p_0} -
e^\theta\right) = 1,\label{8je14}
\end{equation}
where we use the fact that in (\ref{8je10}), $\mu \approx 1$ as can
be
easily deduced.\\
A number of conclusions can be drawn from (\ref{8je14}). For
example, if,
$$A \approx 1, i.e.,$$
\begin{equation}
p_0 \approx \frac{(4\pi )^{5/2}}{1+e}\label{8je15}
\end{equation}
where $A$ is given in (\ref{8je14}), then $z' \approx 1$.
Remembering that in (\ref{8je10}), $\lambda$ is of the order of the
de Broglie wave length and $v$ is the average volume occupied per
particle, this means that the gas gets very densely packed for
momenta given by (\ref{8je15}). Infact for a Bose gas, as is well
known, this is the condition for Bose-Einstein condensation at the
level
$p = 0$ (cf.ref.\cite{huang}).\\
On the other hand, if,
$$A \approx 0 (\mbox{that}\quad \mbox{is}\quad \frac{(4\pi )^{5/2}}{e} \approx
p_0)$$
then $z' \approx 0$. That is, the gas becomes dilute, or $V$ increases.\\
More generally, equation (\ref{8je14}) also puts a restriction on
the energy (or momentum), because $z' > 0$, viz.,
$$A > 0(i.e.p_0 < \frac{(4\pi )^{5/2}}{e})$$
$$\mbox{But \quad if}A < 0, (i.e.p_0 > \frac{(4\pi )^{5/2}}{e})$$
then there is an apparent contradiction.\\
The contradiction disappears if we realize that $A \approx 0$, or
\begin{equation}
p_0 = \frac{(4\pi )^{5/2}}{e}\label{8je16}
\end{equation}
(corresponding to a temperature given by $KT = \frac{p^2_0}{2m}$) is
a threshold momentum (phase transition). For momenta greater than
the threshold given by (\ref{8je16}), the collection of Fermions
behaves like Bosons. In this case, the occupation number is given by
$$\bar n_p = \frac{1}{z^{-1}e^{bE_p}-1},$$
instead of (\ref{8je9}), and the right side equation of
(\ref{8je14}) would be given by $' -1'$ instead of $+1$, so that
there would be no contradiction. Thus in this case there is an
anomalous behaviour of the Fermions.\\
We could consider a similar situation for Bosons also where an
equation like (\ref{8je11}) holds. In this case we have equations
like (\ref{8je15}) and (\ref{8je16}):
\begin{equation}
p_0 \approx \frac{(4\pi )^{5/2}}{1.4e-1}\label{8je17}
\end{equation}
\begin{equation}
p_0 \approx \frac{(4\pi )^{5/2}}{e}\label{8je18}
\end{equation}
(\ref{8je18}) is the same as (\ref{8je16}), quite expectedly. It
gives the divide between the Fermionic and Bosonic behaviour in the
spirit of the earlier remarks. At the momentum given by
(\ref{8je17}) we have a densely packed Boson gas rather as in the
case of Bose Einstein condensation. On the other hand at the
momentum given by (\ref{8je18}) we have infinite dilution, while at
lower
momenta than in (\ref{8je18}) there is an anomalous Fermionisation.\\
Finally it may be pointed out that at very high temperatures,
Fermionisation can be expected, as indeed has been shown
elsewhere\cite{bgsastr}. In any case at these very high
temperatures, we approach the Classical Maxwell Boltzmann situation.
\section{Fractal Statistics}
To sum up Fermions and Bosons are divided into two different
compartments, obeying Fermi-Dirac and Bose-Einstein statistics
respectively. While this is true in general, there are special
situations, for example at very low temperatures or in low
dimensions where the distinction gets some what blurred leading to
Bosonization or Semionic effects. Indeed such an anomalous behaviour
is found experimentally in the superfluidity of $He^3$ in contrast
to $He^4$: Though this is sought to be explained in terms of the
conventional BCS theory, the fact is that there are inexplicable
anomalous
features (cf.ref.\cite{schriffer}).\\
We have shown elsewhere the existence of handedness and the blurring
of Fermi-Dirac statistics
in the two and one dimensional cases.\\
It may also be mentioned that very recent experimental results on
carbon nanotubes\cite{delaney,dress,wildoer,odom} exhibit the one
dimensional nature of conduction and behaviour like low temperature
quantum wires thus confirming the results discussed.\\
We will now argue, in the light of the above results that below the
Fermi temperature, the degenerate electron gas obeys a semionic
statistics, that is a statistics in between the Fermi-Dirac and
Bose-Einstein.\\
We have for the energy density $e,$ in case of sub Fermi
temperatures,
\begin{equation}
e \propto \int^{p_F}_o \frac{p^2}{2m} d^3p \propto
T^{2.5}_F\label{8je20}
\end{equation}
where $p_F$ is the Fermi momentum and $T_F$ is the Fermi
temperature. On the other hand, it is known that \cite{meden,reif}
in $n$ dimensions we have,
\begin{equation}
e \propto T^{n+1}_F\label{8je21}
\end{equation}
(For the case $n = 3,$ (\ref{8je21}) is identical to the
Stefan-Boltzmann law). Comparison of (\ref{8je21}) and (\ref{8je20})
shows that the assembly behaves with
the fractal dimensionality $1.5$. So sub Fermi temperatures and fractal dimensionality
go hand in hand.\\
We now come to a consideration of White Dwarf stars and Neutron
stars. In this case it is known that we have a degenerate assembly
of electrons at the temperature $\sim 10^7 K$, which is below the
Fermi temperature which is $\sim 10^{18} K$
(Cf.\cite{huang,ohanianruffini,zeilik}).\\
Let us now apply the earlier result that there are the anomalous
effects at momenta given by (\ref{8je15}). In this case using the
relativistic energy momentum formula, we can conclude that this
corresponds to a temperature of $10^{28} K$. If this is identified
with the Beckenstein temperature (Cf. for example
\cite{ruffinizang}) given by
$$T = \hbar c^3/8 \pi K M G,$$
where $M$ is the mass this mass turns out to be one gram. The
Schwarzchild radius for such a mass would be $\sim 10^{-28}cms$. This is the meaning of the
threshold (\ref{8je15}).\\
To proceed further, let us consider a Neutron Star. In fact for a
Neutron Star, the dimensionality can be shown to be $2.5$. This is
because, as is known, there are $n \sim 10^{31}$ neutrons per $cc$
in this case \cite{zeilik}. With a typical radius $\sim 10^6 cm$, we
have,
$$R = N^{ \frac{1}{2.5}}l,$$
where $l$ is a typical neutron Compton wavelength. Also the Fermi
and ordinary temperatures are known to be,
$$T^{(N)}_F 10^{11} \, \mbox{and}\, T^{(N)} \sim 10^7$$
So we are below the Fermi temperature.\\
Instead of the usual $e \propto T^4$ we have now $e_N \propto
T^{(\frac{1}{D}+1)} (= T^{3.5})$ that is $e_N < e$ so that energy
$(e-e_N)$ is theoretically available eg. if a Neutron star
gravitationally collapses this energy is available. A White Dwarf
(Cf. \cite{huang}) shines because of gravitational contraction
that takes place slowly.\\
Let us now consider
$$N \sim 10^{23}$$
Also, in this case,
$$kT_F = \frac{\hbar^2}{2m} \cdot \frac{N^{2/3}}{R^2} \sim
\frac{100}{R^2}$$ And
$$R = N^m, \mbox{so \, that}\, R <<1 \, \mbox{if}\, N \sim 10^{23}$$
So
$$kT_F > \frac{100}{R^2} >> 100$$
If
$$\left(\frac{1}{m}\right) \sim 2.5 \, \mbox{as \, for \, a \, Neutron \,
star \, then}\, R \sim 10^{-4}$$ So a cc. of Hydrogen atoms that
contracts to a radius of $10^{-4}cm$, shows up as a mini Neutron
star. Similarly, if $N \sim 10^{26}$, then $R$ would be $10^{-3}cm$
and so on. In principle if such a contraction could be induced, we
would have a mini Neutron star, with the release of energy from the
difference $e - e_N$, seen above.\\
For Black Holes on the other hand, we have shown elsewhere that $D = 2$ \cite{tduniv}.\\

\end{document}